\documentclass{aa}  

\usepackage{graphicx}
\usepackage{txfonts}
\usepackage{lipsum}
\usepackage{subcaption} 
\usepackage{lscape}             % to rotate a single page table, example in appendix.
\usepackage{placeins}    % useful with \FloatBarrier, to keep 

\usepackage{booktabs}
\usepackage[usenames,dvipsnames]{xcolor}
\usepackage[normalem]{ulem}
\usepackage{enumitem}
\usepackage{url}
\usepackage[colorlinks=true,
    linkcolor=blue,
    filecolor=magenta,      
    urlcolor=blue,
    citecolor=blue]{hyperref}

\usepackage{xcolor}
\usepackage{ulem}

\usepackage{caption}
\newcommand*{\Msun}{\ensuremath{{\rm M}_{\odot}}}
\newcommand*{\Rsun}{R$_\odot$}
\newcommand*{\Lsun}{L$_\odot$}
\newcommand{\lsun}{\ifmmode L_{\odot} \else L$_{\odot}$\fi}
\newcommand*{\MESA}{\texttt{MESA}}

\def\msun{\ifmmode M_{\odot} \else M$_{\odot}$\fi}
\def\msunyr{\ifmmode M_{\odot} {\rm yr}^{-1} \else M$_{\odot}$ yr$^{-1}$\fi}
\def\zsun{\ifmmode Z_{\odot} \else Z$_{\odot}$\fi}
\newcommand{\mstar}{\ifmmode M_\star \else $M_\star$\fi}
\newcommand{\tgri}{\ifmmode t_{\rm GRI} \else $t_{\rm GRI}$\fi}
\newcommand{\mgri}{\ifmmode M_{\rm GRI} \else $M_{\rm GRI}$\fi}
\newcommand{\lgri}{\ifmmode L_{\rm GRI} \else $L_{\rm GRI}$\fi}
\newcommand{\macc}{\ifmmode \dot{M}_{\rm acc}  \else $\dot{M}_{\rm acc}$\fi}

\begin{document}

%%%%%%%%%%%%%%%%%%%%%%%%%%%%%%%%%%%%%%%%
% if you use custom commands in your title,
% ensure to check your title when submitting!
%%%%%%%%%%%%%%%%%%%%%%%%%%%%%%%%%%%%%%%%
    \title{A quasi-star is born: formation and evolution of accreting quasi-stars as a pathway to Little Red Dots at non-zero metallicity}

   \titlerunning{Formation and evolution of accreting QSs at non-zero metallicities as a pathway to LRDs} 
   \authorrunning{J.Roman-Garza et al.}
   %\subtitle{Subtitle}

%%%%%%%%%%%%%%%%%%%%%%%%%%%%%%%%%%%%%%%%
% Please separate each author with the \and command
%
% Please do not include ORCIDs next to author names.
% Only ORCIDs authenticated by individual authors in EDPS
% editorial system will be taken into account.
% ORCIDs included here will be removed.
%%%%%%%%%%%%%%%%%%%%%%%%%%%%%%%%%%%%%%%%

   \author{J. Roman-Garza\inst{1,2}\fnmsep\thanks{Corresponding author: jaime.romangarza@unige.ch}
        \and C. Charbonnel\inst{1,3}
        \and D. Schaerer\inst{1,3}
        \and T. Fragos\inst{1,2} 
        \and E. Cenci\inst{1}
        \and R. Marques-Chaves\inst{1}
        \and P. A. Oesch\inst{1,4}
        \and M. Xiao\inst{1}
        }

   \institute{Département d’Astronomie, Université de Genève, Chemin Pegasi 51, CH-1290 Versoix, Switzerland 
         \and
         Gravitational Wave Science Center (GWSC), Université de Genève, CH-1211 Geneva, Switzerland. 
         \and CNRS, IRAP, 14 Avenue E. Belin, 31400 Toulouse, France
         \and Cosmic Dawn Center (DAWN), Niels Bohr Institute, University of Copenhagen, Jagtvej 128, K\o benhavn N, DK-2200, Denmark
}

   \date{Received date; Accepted date}

% \abstract{}{}{}{}{}
% 5 {} token are mandatory

%ds Use this format, which is more compact
\abstract{
% Context
The recently discovered so-called Little Red Dots identified by the James Webb Space Telescope represent a population of compact high-redshift sources whose observed properties have motivated models involving black holes embedded within optically thick gaseous envelopes.
}{
% Aims
To investigate the rest-frame optical emission of Little Red Dots, we model the formation and evolution of quasi-stars, i.e. stellar envelopes supported by the accretion luminosity onto a central black hole, originating from rapidly accreting proto-stars reaching the supermassive star regime ($>10^4$~\Msun) and undergoing general relativistic instability.
}{
% Methods
We compute stellar evolution models with net mass gain rates $=0.01$, 0.1, and 1~\Msun/yr and metallicities $Z=0$--0.01. For the mass gain rates $\ge 0.1$~\Msun/yr, stars remain nearly fully convective with $T_\mathrm{eff}\sim4000$--9000~K.
}{
% Results
The general relativistic instability leading to central BH formation occurs at $M_\star\sim3.5\times10^4$ \Msun\ ($6.8\times10^4$~\Msun) for $\macc=0.1$ \Msun/yr (1 \Msun/yr), at luminosities $L \sim 10^9$~\Lsun. The maximum lifetime of quasi-stars is estimated to be $10^7$--$10^8$~yr, under the assumption that the central black hole is capable of sustaining the surrounding envelope in a state of hydrostatic equilibrium until the envelope has been fully accreted (i.e., $M_{\rm BH,max}/M_{\rm QS}=1$). This estimate is $\sim$100--1000 times longer than the quasi-star progenitor's lifetime.
In an environment that allows for rapid accretion independent of metallicity, the formation, evolution, and properties of quasi-stars are found to be as well independent of metallicity. 
}{
%Conclustions
Comparing the luminosities of our models with those of Little Red Dots at $z<4.5$ ($L_\mathrm{bol}\sim10^{9.5}$--$10^{11.5}$~\Lsun) yields quasi-star masses $10^{4.5}$--$10^{6.5}$~\Msun.
The observed minimum luminosity of $\sim10^{9.5}$~\Lsun\ implies accretion rates $\gtrsim0.1$~\Msun/yr for Little Red Dots progenitors. 
Our models offer a framework supporting quasi-stars as the source of Little Red Dot optical emission, and provide insights into their lifetimes, composition, progenitor's environment as well on their minimum and maximum observed luminosities.
}

   \keywords{
    stars: black holes --
    stars: massive --
    galaxies: quasars --
    black hole physics
    }
   \maketitle
   \nolinenumbers

%%%%%%%%%%%%%%%%%%%%%%%%%%%%%%%%%%%%%%%%%%%%%%%%%%%%%%%%%%%%%%
%\vspace*{-0.4cm}
\section{Introduction}

Since the recent discovery of so-called Little Red Dots \citep[LRDs;][]{Matthee2024Little-Red-Dots} their nature is being debated. While initially thought to be dusty active galactic nuclei (AGN) or massive galaxies \citep[e.g.][]{Kocevski2023Hidden-Little-M,Labbe2023A-population-of}, the absence of classical AGN features, the finding of Balmer breaks exceeding predictions of standard stellar populations, and other characteristics, has led to alternative explanations invoking, for example, gas-enshrouded AGN \citep{Inayoshi2025Extremely-Dense}, direct collapse black holes \citep[DCBH;][]{2026arXiv260114368P}, so-called BH stars \citep[BH$\star$s;][]{naidu2025black,de-Graaff2025Little-Red-Dots}, quasi-stars \citep[QS;][]{Begelman2026Little-Red-Dots}, primordial (PopIII) supermassive stars \citep[SMS;][]{2026ApJ...998..124N,2026arXiv260215935C}, self-gravitating disks accreting onto SMS \citep[][]{2025arXiv250722014Z}, and tidal disruption events in runaway-collapsing clusters \citep[][]{2025ApJ...984L..55B}.

While most of these explanations have in common the co-existence of a BH enshrouded by an optically thick gas envelope, several models are agnostic to their respective formation channel (e.g.~the BH$\star$ scenario), or the proposed formation and evolutionary scenarios differ significantly. Furthermore, DCBH and Pop III SMS models require metal-free gas conditions, which seem difficult to reconcile with the ubiquitous presence of metal-lines in LRDs \citep[e.g.][]{DEugenio2025Irony-at-z6.68:,Perez-Gonzalez2026Little-Red-Dots}, although small amounts of metals could be produced shortly after the formation of the DCBH \citep[e.g.][]{2026arXiv260114368P}. It is therefore important to better understand the formation of these objects, and find metallicity-independent scenarios.

\begin{figure*}
   \centering
    \includegraphics[height = 3.2in]{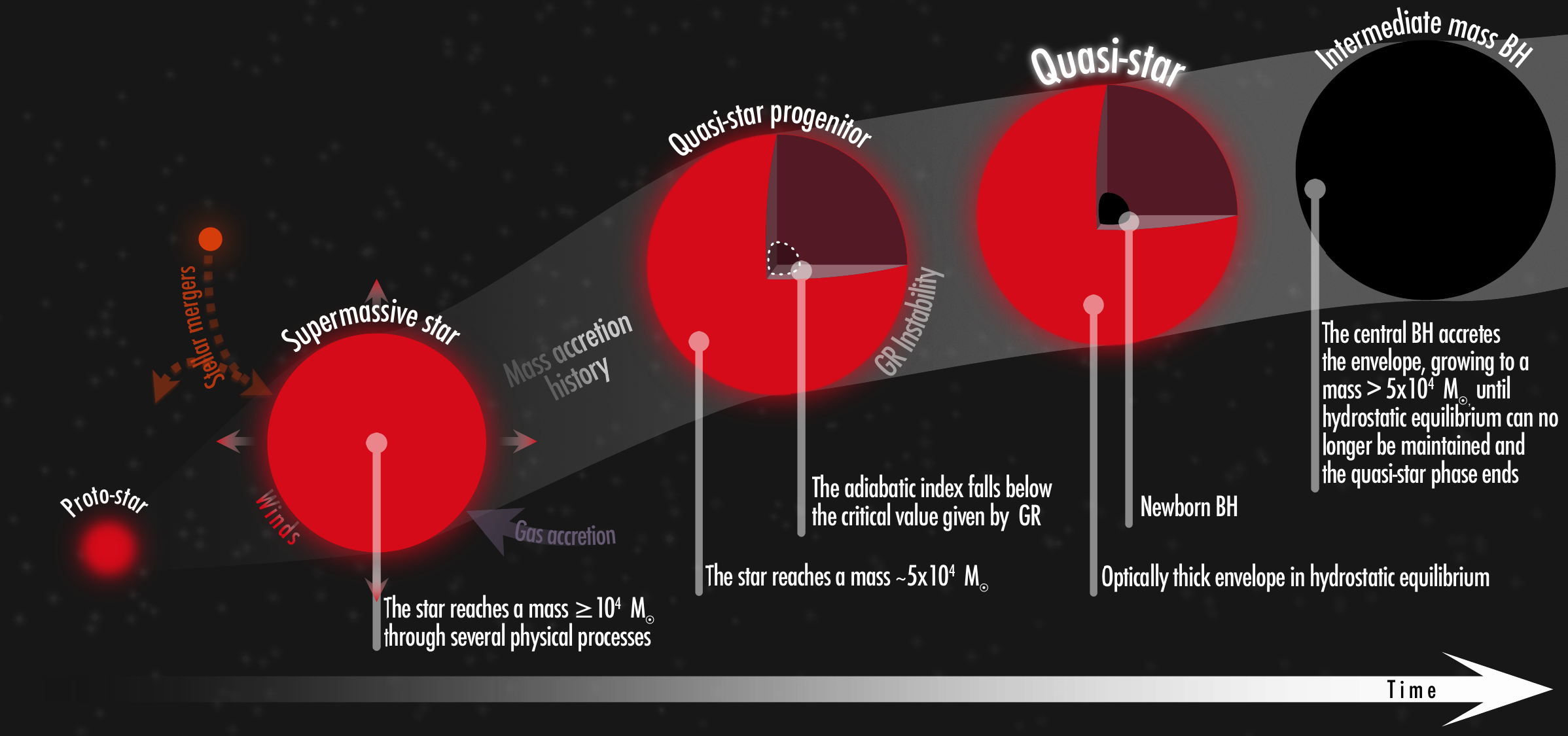} 
    \vspace{0.5cm}
   \caption{Evolutionary stages of the QS and its progenitor considered in the model adopted in this work. The relative sizes of the objects schematically indicate their masses, while the horizontal axis denotes time, both sizes and time are not to scale. From left to right: an initially low-mass proto-star increases its mass through the combined effect of mass-gain and mass-loss processes, including stellar mergers, gas accretion, and stellar winds \citep[e.g.][]{2018MNRAS.478.2461G,2025A&A...699A.223R,2024MNRAS.531.3770R, 2026arXiv260107917R, 2026A&A...707A.163R}. The object evolves into a SMS, attaining a mass $\ge 10^4$~\Msun, and continues to accrete mass. When the SMS reaches a mass of $\sim 5\times 10^5$~\Msun, the adiabatic index, $\Gamma_1$, falls below the critical value predicted by general relativity and the GRI is triggered \citep[e.g.][]{nagele2022stability,2016ApJ...830L..34U,herrington2023modelling}. A central portion of the supermassive star then collapses to form a newborn BH which, depending on its mass, can support the remaining gaseous envelope in hydrostatic equilibrium, transforming the entire system into a QS \citep[e.g.][]{begelman2006formation}. The BH keeps growing in mass by accreting the outer material until hydrostatic equilibrium in the envelope can no longer be sustained \citep[e.g.][]{2011MNRAS.414.2751B,2012MNRAS.421.2713B}; at this stage, the QS reaches the end of its life and is expected to leave behind an intermediate mass black hole (IMBH) as its remnant.
   }
   \label{fig:evolution_diagram}
%\vspace*{-1cm}
\end{figure*}

%\vspace*{-0.1cm}
In this work we explore a metallicity independent scenario where initially low-mass proto-stars experience fast mass gain, e.g. from a net effect of mass loss and gain processes such as stellar winds, gas accretion, runaway stellar collisions in a dense and compact environment; leading to the formation of SMSs ($>10^4$~\Msun) consistent with the literature \citep[e.g.][]{2018MNRAS.478.2461G,2025A&A...699A.223R,2024MNRAS.531.3770R, 2026arXiv260107917R}. Subsequently, the SMS undergo the general-relativistic instability (GRI), leading to the formation of a central BH, becoming quasi-stars, i.e.~objects in hydrostatic equilibrium with a central BH accreting the stellar envelope   \citep[][]{begelman2008quasi}. Fig.~\ref{fig:evolution_diagram} presents a  schematic summary of the model described above in terms of the relevant evolutionary phases encountered by the QS and its progenitor.

Several studies have shown that QS or SMS can reproduce several key observations features of LRDs \cite[see][]{Martins2020Spectral-proper,2026arXiv260512141M,santarelli2026evolutionary,Begelman2026Little-Red-Dots,Sneppen2026Inside-the-coco}. 
Our model "unifies" SMS and QS scenarios for metal-free systems proposed earlier, generalizes them over a broad metallicity range, and proposes a consistent scenario for their formation and evolution, which enforces the possibility that LRDs host QSs.

%%%%%%%%%%%%%%%%%%%%%%%%%%%%%%%%%%%%%%%%%%%%%%%%%%%%%%%%%%%%%%

\vspace*{-0.5cm}
\section{Methods}
\label{s_methods}

To compute the formation and evolution of QS through rapid mass gain onto initially low mass stellar objects, we use the \MESA\footnote{The \MESA\ release 25.10.1 is used for this work, 
as well as \texttt{MESA SDK} 24.7.1.} stellar evolution code \citep{mesa1,mesa2,mesa3,mesa4,mesa5,mesa6}. All models start as a 2~\Msun\  proto-star with an initial radius of $\sim200$~\Rsun, constant entropy, and central temperature of $\sim 10^5$~K. We compute models with metallicities $Z = 0$, $10^{-4}$, $10^{-3}$ and $10^{-2}$ and helium mass fraction $Y = 0.25$, corresponding to [Fe/H]$= -\infty,\ -2.10,\ -1.15\ {\rm and}\ -0.15$ respectively. 

For this work we explore the formation of QS progenitors that maintain a constant mass gain rate independently of their metallicity during most of their lifetime. Previous studies have explored the relevance on the mass loss and gain processes on the formation of EMSs and SMSs, achieving the formation of such objects in the limits of their physical assumptions, considered environments and numerical prescriptions \citep[e.g.][]{2013ApJ...778..178H,2018MNRAS.478.2461G,2025A&A...699A.223R,2024MNRAS.531.3770R,2025MNRAS.539.2561C,2026A&A...707A.163R,2026arXiv260107917R,2026ApJ...999..110N}. Our models gain mass through cold accretion independently of metallicity to explore the formation of QS progenitors across different environment conditions. We consider that such mass gain rate as the net effect of mass gain and loss processes such as stellar mergers, stellar winds and gas accretion. Initially, for numerical stability, we employ the variable accretion rate prescription by \cite{haemmerle2019stellar}, which depends on the bolometric luminosity, increased by a hundredfold. Once the mass accretion rate has reached $\dot{M}_{\rm gain,max}=$ 0.01, 0.1 or 1 \Msun/yr it is kept constant thereafter; this occurs while reaching masses from $\sim 10$--$10^3$~\Msun ~ depending on the threshold value. 

We employed a custom nuclear reaction network to follow nuclear energy generation and compositional evolution. This network extends the \texttt{sagb\_NeNa\_MgAl} network from \MESA, increasing the number of isotopes from 29 to 45. It includes the full set of reactions associated with the CNO cycle, the NeNa and MgAl chains, as well as additional branching channels originating from $^{26}$Al, extending the reaction flow toward $^{27}$Si and $^{28}$Si, and subsequently to $^{29}$P and $^{29}$Si. All these elements are subject to proton-captures during central H-burning in stars of the EMS and SMS mass range \citep[][]{2017A&A...608A..28P,2025A&A...699A.223R}.

To account for general relativistic effects, the gravitational constant is modified according to the first order correction given by the Tolman–Oppenheimer–Volkoff equation, which is derived for the Schwarzschild metric. The correction is computed as \citep[e.g. see][]{herrington2023modelling}:
    \begin{equation}
    G_{\rm TOV} = G 
    \left( 1 + \frac{P}{\rho c^2} + \frac{4 \pi P r^3}{m_r c^2} \right)
    \left( 1 - \frac{2 G m_r}{r c^2} \right)^{-1} .
    \end{equation}
    
    The evolution of the stars stops when the general-relativistic instability (GRI) criterion by \cite{nagele2022stability} is reached:
    %i.e., if
    \begin{equation}
        \Gamma_{\rm 1}(m_r) \le \frac{4}{3} + 4.498\frac{ G_{\rm TOV} m_{r}}{r c^2}f_{\rm GR}.
        \label{eq:gri_criterion}
    \end{equation}
    $\Gamma_{\rm 1}(m_r)$ corresponds to the value of the adiabatic index at the mass coordinate $m_{r}$ , which corresponds to the radial coordinate $r$. The factor $f_{\rm GR}$ is introduced and chosen as 1.5 to stop the evolution of the stellar models before they reach the GRI and become numerically unstable. The models stop when they have $\gtrsim70\%$ of the stellar mass needed to reach the GRI given $f_{\rm GR} = 1$. 
    
    When the GRI is encountered (i.e., Eq.~\ref{eq:gri_criterion} is satisfied) the stellar evolution run is halted. The core of the corresponding supermassive star is expected to collapse into a BH \citep[][]{loeb1994collapse,volonteri2005rapid,begelman2006formation}. It then enters a new evolutionary phase as a QS \citep[][]{begelman2008quasi}, where the system is stabilized against gravity primarily by the radiation produced from accretion onto the central BH. Stars that encounter the GRI after the main sequence may experience the general relativistic instability supernova (GRISN) where no BH is expected to be formed \citep[see][]{nagele2022stability,haemmerle2018evolution}. For this work we do not explore the cases where the GRISN is triggered;  with the adopted mass growth rates, the stars reach the GRI while on the main sequence.

    The evolution of the QS model is computed with the \texttt{MESA-QUEST} module \citep[see][]{santarelli2026mesa}. The central boundary conditions on mass and luminosity of the QS models are set at the radius of the sphere of influence of the BH ($R_{\rm in}$) that is assumed to be the Bondi radius of the BH. We investigate two limiting cases, one where gas accretion stops as soon as the GRI is encountered, and one where accretion onto the QS continues at the same rate as previously. In both cases, the central boundary condition of the model is set to an initial BH mass of 10~\Msun. The central BH grows in mass through the accretion of the stellar envelope  following the convection-limited Bondi accretion rate \citep[][]{2012MNRAS.421.2713B}, with both convective and radiative efficiencies set to $0.1$. The BH accretion luminosity is set in terms of the radiative efficiency and accretion rate as $\simeq 0.11\ \dot{M}_{\rm acc}\ c^2$ 
    \citep[see][]{santarelli2026mesa}. 
    
    The initial BH mass is chosen to ensure numerical convergence of the simulation rather than being set by a physical assumption. We adopt the same value of 10~\Msun\ as in \cite{santarelli2026evolutionary}, while other QS models assume 100~\Msun\ for the newborn BH mass \citep[see][]{2026ApJ...998...65H}. Concerning numerical results on the initial BH mass, \cite{fuller1986evolution} report that, after the onset of the GRI, about 80\% of the SMS mass initially collapses homologously in non-exploding models, but once neutrino luminosity becomes important, only a small central fraction of the QS continues to collapse in this manner. By the time their simulation is halted, they state that only 10--20\% of the SMS mass is still collapsing homologously; however, their data \citep[see Fig.~9 in][]{fuller1986evolution} indicate that this fraction may be smaller, so 10\% is interpreted as an upper limit on the newborn BH mass. Subsequent SMS collapse studies \citep[e.g.][]{2017PhRvD..96d3006S,2012ApJ...749...37M} typically employ an $n=3$ polytropic SMS, implying that the entire QS progenitor lies in the Newtonian instability regime and it is already beyond the GRI threshold, so that $\gtrsim 90\%$ of the SMS mass collapses into a BH. Finally, \cite{2012ApJ...749...37M} compare $n=3$ polytropes with initially hydrostatic SMS models, finding that only the former collapse. Unfortunately, the hydrostatic models are not evolved up to the GRI, preventing a direct comparison with \cite{fuller1986evolution}. Given the previous discussion, the mass of the newborn BH remains uncertain.

%%%%%%%%%%%%%%%%%%%%%%%%%%%%%%%%%%%%%%%%%%%%%%%%%%%%%%%%%%%%%%
\vspace*{-0.4cm}
\section{Results}
\label{s_results}

\subsection{Overview of the evolution from proto-star to quasi-star}
\label{sec:overview}

\begin{figure}
   \centering
    \includegraphics[width=\columnwidth]{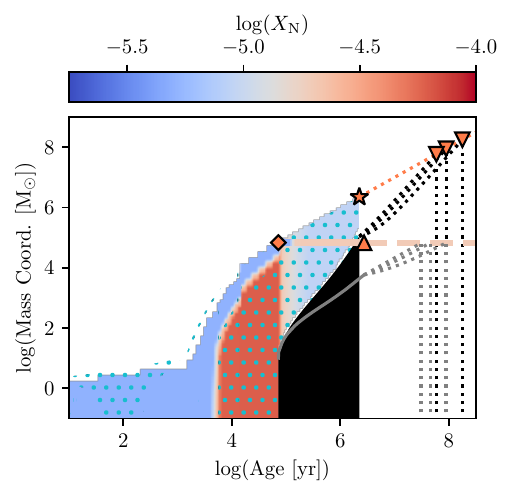} 
   \caption{Kippenhahn diagram showing the mass coordinate as a function of time for the model with $\dot{M}_{\rm gain,max} = 1$~\Msun/yr, $Z=10^{-4}$ and $Y=0.25$. The nitrogen mass fraction is color-coded; 
   convection zones are indicated by cyan dots. The diamond shows the instant of GRI. The black filled region shows the BH mass. The non-accreting QS model is shown by the thick dashed line, the color of which corresponds to the surface N abundance as a function of time, and the corresponding BH mass by the solid gray line. The extrapolated QS mass for the accreting model is shown by the dotted orange line; that of the non-accreting model remains constant (orange dashed). The extrapolated BH masses are plotted with black and gray dotted lines, respectively; in each case, the three lines correspond to different estimates assuming $\alpha = 0.5,\ 1,\ {\rm and}\ 1.5$. The star and  the upward triangle indicate the crashing points of the accreting and non-accreting QS models. The downward triangles indicate the extrapolated masses of the object when the envelope has been entirely swallowed by the BH. 
   }
   \label{fig:kipp_1}
%\vspace*{-1cm}
\end{figure}

%%%%%%%%%%%%%%
The evolution of the accreting object through the proto-star, SMS, and QS phases of the model with the highest accretion rate ($\dot{M}_{\rm gain,max} = 1$~\Msun/yr) and $Z= 10^{-4}$ is shown through a Kippenhahn diagram in Fig.~\ref{fig:kipp_1}.
After $\tgri \sim 7 \times 10^4$ yr and for a mass of $\mgri \sim 6.8\times10^4$~\Msun\ the object reaches the GRI and enters the QS phase. This occurs while the object is still burning hydrogen early on the main sequence (the central He mass fraction has reached $\sim 0.30$ only). 
The QS evolution then depends on the accretion rate of the central BH, $\dot{M}_{\rm BH}$, as well on whether the QS continues to accrete material and on the maximum mass the BH can reach while maintaining a hydrostatic envelope.

The maximum mass the central BH can reach depends on the size of the BH's sphere of influence, $R_{\rm in}$, as well as on the accretion history of the QS and the BH accretion rate. Regarding the former point, we recall  that we consider $R_{\rm in}$ to be equal to the BH's Bondi radius, as first proposed by \cite{2011MNRAS.414.2751B}, where it is presented as a physically motivated assumption, although the size of this region is uncertain. For the same assumption on $R_{\rm in}$, \cite{2012MNRAS.421.2713B} found that no QS can exist if $M_{\rm BH}>1\%\, M_{\rm QS}$, as no hydrostatic solutions are obtained for such BH masses. Later, \cite{2024ApJ...970..158C} explained the dependence of the maximum BH mass on $R_{\rm in}$, showing that a larger value than the one assumed by \cite{2011MNRAS.414.2751B} lead to a higher limiting BH mass, obtaining QS models with $M_{\rm BH}\simeq 60\%\, M_{\rm QS}$. More recent studies have produced QS models with maximum BH masses ranging from $\sim10\%\, M_{\rm QS}$ \citep[][]{2026ApJ...998...65H} up to $\sim50\%\, M_{\rm QS}$ \citep[][]{santarelli2026evolutionary}.

Consistently with these studies, our simulations manage to reach the point where $M_{\rm BH}\sim10\%$ of the total QS mass at the current epoch, where \MESA\ encounter convergence issues as the remaining envelope is marginally bound with a mass average value of $\Gamma_1$ only $0.05\%$ above the classical instability limit of $4/3$. Convergence issues are also encountered due to a density inversion  near the surface of the QS envelope. 

\begin{figure}
   \centering
    \includegraphics[width=3.5in]{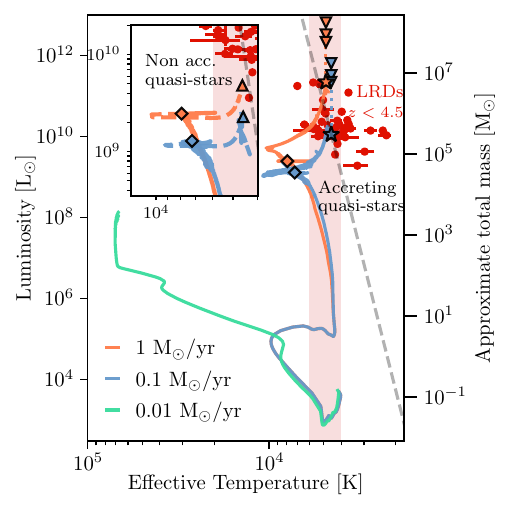} 
   \caption{Evolution in the HRD of the $Z=10^{-4}$ models with the accretion rates color-coded. The approximate total mass of the star, or QS, is given by the right vertical axis (estimated by Eq.~\ref{eq:ledd}). The solid lines represent the continuously accreting models. The non-accreting QSs are shown in dashed lines inside the nested plot. The diamonds, star, and triangles  correspond to the same phases as in Fig.\ref{fig:kipp_1}. The maximum values of the luminosity for the accreting QSs 
   are estimated by the extrapolation of the QS mass and considering
   a constant $T_{\rm eff}$. The gray dashed line is a fit to the ending points of the QS models by \cite{santarelli2026evolutionary}. The red shaded region corresponds to the effective temperature range of the Hayashi line as considered by \cite{inayoshi2025critical}. The observed LRD sample from \cite{de-Graaff2025Little-Red-Dots} is shown with red dots. }
   \label{fig:HRD}
\end{figure}

Given the uncertainty on the maximum BH mass for a QS, as described above, and its impact on the total lifetime of the object we analytically extrapolate the BH mass after the \MESA\ simulations stop until the point the BH has accreted all the envelope (see Appendix~\ref{app:lifetime_qs}).  To obtain an upper limit of the BH mass as well as the QS lifetime, we integrate Eq.~\ref{eq:mdot_bh} over time, considering the values of the free parameter $\alpha = 0.5, 1\ {\rm and }\ 1.5$ that accounts for possible enhancements or decrements of the BH accretion rate by, e.g., magnetic fields or angular momentum transport. Assuming $\dot{M}_{\rm gain,max} = 1~\Msun/\mathrm{yr}$ and that accretion is halted at the GRI, the estimated maximum QS lifetime is $\lesssim 10^8~\mathrm{yr}$, and the final mass of the QS therefore provides an upper limit to the final BH mass. In this case, the BH can reach a maximum mass of $6.8\times 10^4$~\Msun\ by the end of the QS phase, that belongs in the intermediate mass range of BHs. If accretion onto the QS is maintained, the BH can grow up to $\sim 10^8$~\Msun ~within $\sim 10^8~\mathrm{yr}$, given the adopted mass accretion rate. Note that the estimated QS lifetime is considered an upper limit, as our analytical results underestimate how $\dot{M}_{\rm BH}$ grows in time by comparing results from the simulations and the analytical formula. The different estimates given the value of $\alpha$ are also considered to explore this uncertainty. Moreover, if the QS phase terminates before the BH has fully accreted the envelope, the object's lifetime can be approximated as our earlier estimate multiplied by the factor $M_{\rm BH} / M_{\rm QS}$ (see Appendix~\ref{app:lifetime_qs}). The exact point at which the QS phase terminates remains uncertain, as studies suggest it occurs when $M_{\rm BH} / M_{\rm QS}$ lies between 0.01 and 0.6 \citep[see][]{2011MNRAS.414.2751B,2012MNRAS.421.2713B,2024ApJ...970..158C,santarelli2026evolutionary,2026ApJ...998...65H}. We remind that for this reason, we focus on the maximum lifetime estimates.

For models with $\dot{M}_{\rm gain,max} = 0.1~\Msun/\mathrm{yr}$, the GRI occurs somewhat later ($\tgri \sim 0.349 $ Myr) and at a lower mass ($\mgri \sim 3.5 \times 10^4$ \msun), but the maximum QS lifetime is similar to that of models with $\dot{M}_{\rm gain,max} = 1~\Msun/\mathrm{yr}$. The main estimation is that QSs are expected to last no more than $\simeq 10^7$ to $10^8~\mathrm{yr}$ (see Appendix~\ref{app:lifetime_qs}). For comparison, the lifetime of the non accreting QS models from \cite{santarelli2026evolutionary}, that attain a final value of $M_{\rm BH}/M_{\rm QS}\simeq 0.5$, is estimated to be 30--40~Myr, that is consistent with our results. The expected lifetime of the QS progenitors  depends on the mass gain rate but, as an order of magnitude approximation,  it is 100--1000 times shorter than QS lifetime. For a more quantitative comparison between all QS models, see Appendix~\ref{app:relevant_quant} and the table therein.
%ds
Note also that the case with $\dot{M}_{\rm gain,max}=0.01$ \Msun/yr did not reach the GRI and was stopped earlier for numerical reasons.

We note that the assumption on the initial BH mass could lead to an overestimation of the QS lifetime. As shown in Fig.~\ref{fig:kipp_1}, if the true initial BH mass is in the order of $\sim 100$~\Msun\ then the QS lifetime can be overestimated by about $\lesssim 0.1$~Myr, that represent at most 1\% of our upper limit estimates of the QS lifetime. In another case, if the true newborn BH mass is $\sim1000$~\Msun, corresponding to $\sim10\%$ of the QS mass at $t_{\rm GRI}$, then the QS lifetime will be overestimated on the order of $\lesssim 1$~Myr. 

    \begin{figure}
           \centering
            \includegraphics[width=3.5in]{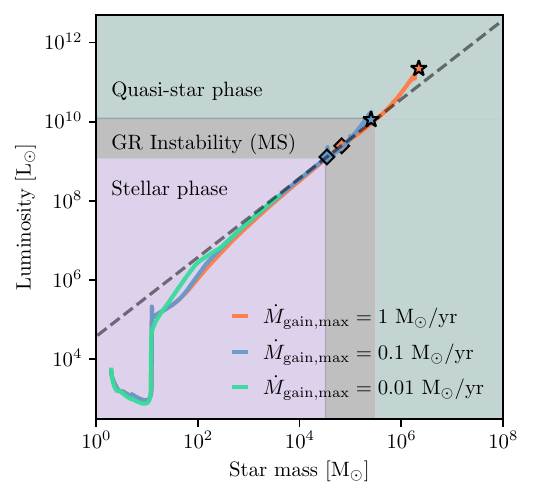} 
           \caption{Luminosity as a function of stellar mass for the continuously accreting models: 1~\Msun/yr (orange), 0.1~\Msun/yr (blue), and 0.01~\Msun/yr (mint). The instant when the GRI criterion is met is marked by diamond symbols in the corresponding colours. The grey shaded region denotes the mass range where the GRI is expected to occur, for main-sequence stars with accretion rates from 0.1 to 1~\Msun/yr \citep[see][]{2016ApJ...830L..34U,2017ApJ...842L...6W,2013ApJ...778..178H,haemmerle2018evolution,nagele2022stability,herrington2023modelling}, and their corresponding luminosities. The black dashed line represents the mass–luminosity relation obtained by setting the stellar luminosity equal to the Eddington luminosity.}
           \label{fig:mass_luminosity}
        \end{figure}

\subsection{Nucleosynthesis and chemical properties}
The nucleosynthesis and the expected variations of the surface chemical properties of rapidly accreting SMS and QS are also worth  discussing. 
Before reaching the GRI, the star burns H through the CNO cycle as well as the NeNa and MgAl chains, in its very massive convective core. 
No further nucleosynthesis occurs during the QS phase, since the temperature and density of the envelope are too low.
They can thus synthesize a maximum mass of nitrogen  
$M_N = \overline{X_N} \mgri \le X^{\rm ini}_{\rm CNO} \mgri \approx (1-1.5)\times 10^{-2} \, 10^{[{\rm O/H}]^{\rm ini}} \mgri $, where $\overline{X_N}$ is the average nitrogen mass fraction in the star and $X^{\rm ini}_{\rm CNO}$ the initial mass fraction in C+N+O. For the model shown in Fig.~\ref{fig:kipp_1} with $[{\rm O/H}]^{\rm ini}=-2.1$, this corresponds to $M_N \approx (5-8)$ \msun, in agreement with the predicted average mass fraction $X_{\rm N} \sim 10^{-4}$ before the GRI is reached. However, in our SMS model, the very extended convective core is surrounded by a radiative envelope. Therefore, the fresh N is expected to appear at the surface only at the beginning of the QS phase, when the envelope becomes fully convective and mixes up the nucleosynthesis products. Simultaneously to the N enrichment, the surface abundances of C and O are dropping (not shown in the figure) as a result of CNO-cycle.  No more N is produced  later, as nucleosynthesis is quenched in the QS since the temperature and density in its envelope are too low. If this object stops accreting, the surface abundances then remain constant, with a value for N that is depicted by the color of the dashed line in Fig.~\ref{fig:kipp_1}. However, if accretion continues beyond the GRI, the surface abundance of N rapidly decreases again due to dilution with the accreted matter, as shown in Fig.~\ref{fig:kipp_1}. Therefore, the CNO equilibrium values (high N/O and N/C) can only be reached during a short amount of time at the very beginning of the QS phase or if the objects stop accreting close to or before the GRI.  

Additionally, once the central temperature of the accreting star exceeds  respectively $\sim$ 40 and $70$~MK, 
the NeNa and MgAl chains are efficiently triggered \citep[e.g.][]{2017A&A...608A..28P}. As a consequence, Na and Al are produced while Ne and Mg are depleted in the stellar convective core (see fig.\ref{fig:kipp_1_NaAlMg} in Appendix~\ref{app:mgal_abundances}). After the QS forms, the central hot-H burning yields are mixed throughout the remaining, now fully convective, envelope. As previously described for C, O, and N, Ne- and Mg-depleted but Na- and Al-enriched material is brought to the surface.   
Again, the maximum surface abundance variations are kept in the case the QS stops accreting around the occurrence of the GRI, while in the accreting case they are progressively erased by dilution with the further accreted material as the luminosity of the QS increases across the luminosity range of LRDs.

\cite{2026arXiv260215935C} proposed that LRD could be globular clusters (GC) in formation, hosting a short-lived SMS releasing H-burning yields compatible with the abundance variations of GC multiple stellar populations. 
In our SMS-QS scenario, only the less luminous LRDs corresponding to QS that have stopped accreting around the luminosity of the GRI should exhibit persistent abundance signatures of hot-H burning resembling those of proto-GCs.  There, the expected N enrichment is similar to that observed among the second population stars in GCs, as the CNO-cycle was running at equilibrium in the SMS. However, the Na enrichment is relatively limited as the fresh Na coming from p-captures on Ne is partly depleted when the central temperature of the SMS is high enough to deplete Mg \citep[e.g.][]{2017A&A...608A..28P,2018MNRAS.478.2461G}. On the other hand, the accreting QS %that cover the luminosity range of the LRDs 
would show these very specific abundance patterns only during a brief episode at the relatively low luminosity around that where the GRI occurs. These patterns are attenuated and eventually erased as the luminosity of the QS increases towards that of the brightest LRDs.  \\

\subsection{Evolution in the HR diagram}
\label{sec:ev_hrd}

Figure~\ref{fig:HRD} shows the evolutionary paths across the Hertzsprung-Russell diagram (HRD) of our models with $Z=10^{-4}$. Starting from the lower-right corner of the HRD, the initial proto-star first contracts and moves towards higher effective temperatures and luminosities, exiting the domain defined by cool and fully convective stellar envelopes (i.e., $4000~{\rm K} \lesssim T_{\rm eff} \lesssim 6000~{\rm K}$), referred in the literature as the Hayashi line \citep[see][]{hayashi1961stellar,inayoshi2025critical}. When the luminosity reaches $\sim 3\times 10^4$~\Lsun, the evolutionary tracks bifurcate. The model with $\dot{M}_{\rm gain,max} = 0.01$~\Msun/yr departs towards higher effective temperatures (we remind that this model was computed up to a mass of $\simeq 3900$~\Msun\ and did not encounter the GRI). The other two models return towards the Hayashi line. This behavior is expected for  accretion rates above $\sim 0.02$~\Msun/yr, as extensively discussed in the literature \citep[e.g.,][]{2012ApJ...756...93H,2013ApJ...778..178H,herrington2023modelling,2025A&A...699A.223R,nandal2023critical}. Above this value, accretion causes the inflation of the stellar radius, as the timescale to radiate away the advected entropy becomes longer than the Kelvin-Helmholtz timescale. Consequently, models with $\dot{M}_{\rm gain,max} \ga 0.1$ \Msun/yr follow  similar tracks, with the luminosity increasing at relatively low effective temperature as the stellar mass increases.  Above $10^6$~\Lsun, both models gradually depart from the Hayashi line toward higher effective temperatures as they cease to be fully convective, returning once convection again dominates their structure.

\begin{figure}
       \centering
        \includegraphics[width=3.5in]{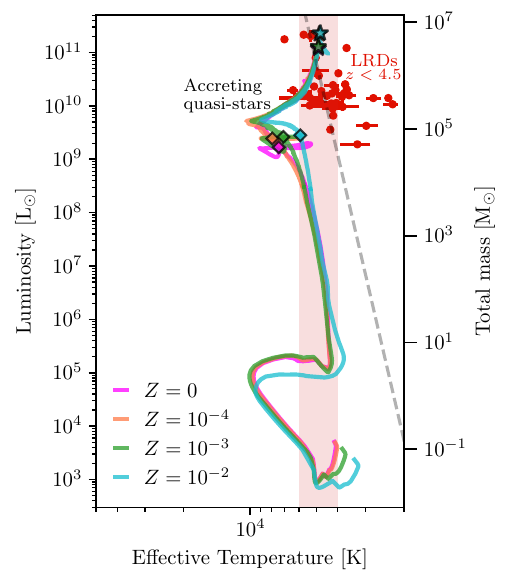} 
       \caption{Similar to Fig.~\ref{fig:HRD}, showing accreting models with $\dot{M}_{\rm gain,max} = 1$~\Msun/yr and different metallicities color-coded.}
       \label{fig:HRD_z}
    \vspace*{-0.4cm}
    \end{figure}

The luminosity of SMSs is proportional to their mass and independent of the accretion rate \citep[see][]{haemmerle2018evolution}. Once the QS is formed, the luminosity of the remaining stellar envelope is regulated primarily by the QS total mass (central BH plus envelope), as previously found for metal-free QSs \citep[e.g.][]{begelman2008quasi,santarelli2026evolutionary}. Fig.~\ref{fig:mass_luminosity} show the mass-luminosity relation for the continuously accreting models. For stars with a mass $\gtrsim 100$~\Msun\ the relation is independent of the accretion rate and metallicity \citep[see e.g.][]{Martins2020Spectral-proper}, and it essentially follows the Eddington luminosity of the object,  
    \begin{equation}
        L_{\text{Edd}} = \frac{4\pi G m_p c}{\sigma_T} M = 3.7 \times 10^4 \lsun \, \left(\frac{M}{\msun}\right).
        \label{eq:ledd}
    \end{equation}

As indicated by Fig.~\ref{fig:mass_luminosity}, the total stellar mass required to reach the GRI depends on the accretion rate \citep[this has been discussed as well by][]{2016ApJ...830L..34U,2017ApJ...842L...6W,2013ApJ...778..178H,haemmerle2018evolution,nagele2022stability,herrington2023modelling}. For accretion rates between 0.1 to 1~\Msun/yr, the stars will encounter the GRI when they reach masses between $\sim 10^{4.5}$ to $\sim 10^{5.5}$~\Msun, and it is expected that they will be in the MS at that point. Once the QS phase begins the relation between mass and luminosity stays similar to the one in the stellar phase \citep[e.g.,][]{santarelli2026evolutionary}.

  At the instant when the GRI is encountered our models have a luminosity of $L_{\rm GRI}(\dot{M}\gtrsim0.1\,{\rm M}_\odot/{\rm yr}) \simeq 10^9$~\Lsun, and they become more luminous as the QS mass increases. For example, the model with $\dot{M}_{\rm gain,max} = 1$~\Msun/yr reaches a total mass of $\sim 2.2\times 10^6$~\Msun\ at the point the simulation stops due to convergence issues, corresponding to a luminosity of $\sim 2.2\times 10^{11}$~\Lsun. If we extrapolate to the end of life of the QS, we find that it can reach a maximum luminosity of almost $10^{13}$~\Lsun\ if it continues to grow in mass. In the case of non-accreting QS, our results are consistent with the models from \cite{santarelli2026evolutionary}. For a summary on the evolution in luminosity, effective temperature, stellar mass, and BH mass across different evolutionary epochs, see Appendix~\ref{app:relevant_quant} and the table therein.

To study the effect of metallicity on the evolution  of accreting QSs we computed additional models with $\dot{M}_{\rm gain,max} = 1$~\Msun/yr at $Z = 0$, 0.001 and 0.01. As shown in Fig.~\ref{fig:HRD_z}, their evolution tracks are very similar, and they overlap remarkably during the QS phase. 
Also, the mass when GRI occurs, \mgri, increases only by a factor of two (from 46000 to 83000~\Msun) from zero to solar metallicity. Nucleosynthesis depends on metallicity, but is largely determined by the composition of accreted matter (see above).

\subsection{LRDs as QS hosts and implications on BH masses} 
\label{sec:lrd_and_qs}

In Fig.~\ref{fig:HRD}, we compare the predicted tracks of our SMS and QSs in the HR diagram with the LRD sample of \cite{de-Graaff2025Little-Red-Dots}, for which bolometric luminosities and effective temperatures were obtained by fitting modified blackbody spectra to the rest-optical part of the observed  spectra. The LRDs in this sample have luminosities $\sim 10^{9-11}$ \Lsun. Their inferred temperatures are comparable to those predicted by our models, especially if uncertainties in the effective temperature are considered \citep[e.g. the efficiency of convection as shown in][]{santarelli2026evolutionary}.

The comparison between the predicted and observed HRD and the position of these objects above $L > 10^9$~\Lsun$\sim L_{\rm GRI}$ may be explained by LRDs hosting QSs, which dominate their rest-optical emission.  Furthermore, since the value of $L_{\rm GRI}\simeq 10^9$~\Lsun\ depends on the accretion rate, it is also noted that QS progenitors should be formed through accretion rates $\gtrsim 0.1$~\Msun/ year.
Our results also imply that LRDs are intimately linked to BHs, as suggested by numerous earlier studies \citep[e.g.,][]{Begelman2026Little-Red-Dots,Ma2025No-Luminous-Lit,Greene2026What-You-See-Is}. The corresponding bolometric luminosity provides a direct measure of the total mass of the QS, and hence an upper limit of the BH mass, since it represents an a priori unknown fraction of the QS mass. For example, predictions by \cite{Greene2026What-You-See-Is}, based on the Eddington limit, may overestimate the BH mass up to two orders of magnitude.

Our scenario also predicts that objects with similarly low temperatures but luminosities $ \lesssim L_{\rm GRI}$ should exist. One possibility is for them to be SMSs, but they should represent, as order of magnitude approximation, $\lesssim 0.1$--$0.01\%$ of the overall population, since the stellar phase is $\sim100$ to $\sim1000$ times shorter than the lifetime in the QS phase, as discussed previously (cf.~Appendix~\ref{app:lifetime_qs}). As an order of magnitude estimation, the ratio between the QS lifetime with respect to the lifetime of its progenitor remains similar for the cases where the newborn BH mass is $< 10\%$ of the QS mass. The other possibility evokes QSs whose progenitors accreted in a rate $<0.1$~\Msun/yr. The lack of LRDs with $L < 10^9$~\Lsun\ may indicate that the progenitors of the inhabiting QSs had accretion histories of $\gtrsim 0.1$~\Msun/yr.

Although we have not been able to compute consistent numerical models of QSs with masses well above $10^6$ \Msun, our analytical estimates show that such objects could grow up to masses of $\sim 10^8$ \Msun\ in case of continuing gas accretion with a rate of 1 \msunyr\ and be dominated by the BH mass (Fig.~\ref{fig:kipp_1} and Table \ref{tab:results}). This would translate to luminosities up to $\simeq 10^{13}$~\Lsun, approximately a factor $\sim 10$ times more luminous than the brightest LRD identified so far \citep[see][]{Ma2025No-Luminous-Lit}. Nonetheless, the existence of the QS also depends on the maximum BH mass that can support the gaseous envelope in hydrostatic equilibrium. No significant rest-frame optical emission is expected for systems in which the BH exceeds this mass, as no envelope will remain to enshroud it. Given our accreting models, a maximum BH mass of $\simeq 10\%$ for a QS with mass of $\sim2\times10^6$~\Msun\ is enough to reproduce the observable properties of the most luminous LRDs in the sample by \cite{de-Graaff2025Little-Red-Dots}. Whether nature indeed forms significantly more luminous LRDs, and whether the scenario indicated here holds, remains to be seen.

%%%%%%%%%%%%%%%%%%%%%%%%%%%%%%%%%%%%%%%%%%%%%%%%%%%%%%%%%%%%%%
\vspace*{-0.4cm}
\section{Conclusions}
\label{s_conclude}

As a pathway to understand the physical origin of the rest-optical emission from LRDs, we have calculated stellar evolution models of rapidly accreting stars. We start from low mass-proto-stars that experience a net mass gain and follow their evolution until they become SMS and reach the GRI, which leads to the formation of so-called QS \citep[where a stellar envelope is supported by a central BH, see][]{begelman2006formation,begelman2008quasi}. We considered maximum mass gain rates of 0.001, 0.1 and 1~\Msun/yr, and metallicities from $Z=0$ to 0.01. We show that the evolutionary tracks of our models are independent of metallicity, as long as the mass gain history remains independent of metallicity.

 We are aware that the employed mass gain rates are in agreement with the mass-inflow rates onto simulated haloes from the cosmological simulations. For example, \citet{cenci2025little} reports inflow rates range from $\sim 10^{-3}$ to $\gtrsim1$~\Msun/yr at $z\simeq4$, which increase with higher redshifts. The mass growth threshold can also be reached in dense and compact forming massive star clusters with stellar surface densities similar to those estimated for LRDs (typically $>10^5$~\Msun\,pc$^{-2})$, where both gas-accretion and collisions can concur to the formation of SMS, with no limitation in terms of metallicity, contrary to the DCBH scenarios \citep[][and references therein]{2018MNRAS.478.2461G,2025MNRAS.543.1023L,2026arXiv260202702B,2025A&A...699A.223R}. 

Accreting models with $\dot{M}> 0.01$~\Msun/yr remain almost fully convective across the entire evolution, with effective temperatures from 4000 to 9000~K~\citep[consistent with][]{herrington2023modelling,haemmerle2018evolution,nandal2023critical}. Stars accreting at 0.1, 1~\Msun/yr and $Z=10^{-4}$ reach the GRI when they have masses of $3.5\times 10^4$ \Msun\ and $6.8\times 10^4$~\Msun\ respectively. Their luminosity at that instant is $\simeq 10^9$~\Lsun. 

Significant amounts of material processed by the CNO cycle (N, in particular) and the NeNa and MgAl chains (Al in particular, and Na to a lower extent) can build up during the SMS phase preceding the GRI. However, the continuing mass growth through the QS phase, which is needed to explain the observed luminosity range of LRDs, implies a strong dilution, and thus variable and not very strong chemical enrichment in QS.

The QS lifetime upper limit is estimated on the order of $10^7 - 10^8$~yr, consistent with the values the numerical models by \cite{santarelli2026evolutionary,2026ApJ...998...65H}. This is approximately 2-3 orders of magnitude longer than the lifetime of their stellar progenitors, depending as well on the progenitor's accretion history. 
Therefore, it is much more likely to observe QSs than their progenitors. Nevertheless, the lifetime of the QS depends on the assumptions of our model, such as the initial BH mass as well as the BH's accretion rate and the size its sphere of influence, that remain uncertain.

Our accreting QS models are in agreement with the effective temperatures and range of bolometric luminosities of observed LRDs, showing $L \sim 10^{9.5-11.5}$ \Lsun\ \citep{de-Graaff2025Little-Red-Dots}. This suggests that LRDs may be dominated by QSs with total masses between $\sim 10^{4.5-6.5}$ \Msun.
The mass of their central BH depends on the evolutionary stage of the QS and its accretion history, and the QS mass serves only as its upper limit. Assuming that the QS mass is of the same order as the BH mass \citep[as in][]{Greene2026What-You-See-Is} may overestimate the value of the latter.

From our models, the minimum bolometric luminosity of $\simeq 10^9$~\Lsun\ of LRDs implies that the QS progenitors should have a minimum accretion rate of $\simeq 0.1$~\Msun/yr. If such minimum observed luminosity was higher it would imply that the minimum accretion rate of the QS progenitors is $>0.1$~\Msun/yr and vice versa. The accreting QS framework may impose an upper bound on its luminosity, corresponding to the value reached when the BH attains the largest mass for which the surrounding envelope can still maintain hydrostatic equilibrium. However, this critical BH mass is still poorly constrained \citep[see][]{2011MNRAS.414.2751B,2024ApJ...970..158C,2026ApJ...998...65H}. If this maximum BH mass is $\sim10\%$, and the QS grows at a steady rate of 1~\Msun/yr, then the peak QS luminosity is consistent with the highest LRD luminosities observed at $z>4.5$.

%%%%%%%%%%%%%%%%%%%%%%%%%

\begin{acknowledgements}
      This work was supported by the Swiss National Science Foundation (PI Fragos, project number CRSII5\_213497). The authors thank M. Habouzit for the discussions regarding this work. As well, we thank the referee for the valuable and constructive input that helped to improve this manuscript. This work has received funding from the Swiss State Secretariat for Education, Research and Innovation under contract number MB22.00072, as well as from the Swiss National Science Foundation through project grant 200020\_207349. The Cosmic Dawn Center is funded by the Danish National Research Foundation under grant DNRF140.
\end{acknowledgements}

\vspace*{-0.4cm}
\bibliographystyle{aa} 
\bibliography{references}

%%%%%%%%%%%%%%%%%%%%%%%%%%%%%%%%%%%%%%%%%%%%%%%%%%%%%%%%%%%%%%
\begin{appendix}

\section{Lifetime of quasi-stars}
\label{app:lifetime_qs}

    The aim of this section is provide an analytical prediction for the lifetime of QSs depending on the accretion history during this phase. We define that the QS progenitors have a constant mass gain rate $\dot{M}$, therefore their mass across time is modeled as 
    \begin{equation}
        M(t) = \dot{M}\,t, \quad {\rm if }\ t< t_{\rm GRI}.
    \end{equation}
    Where $t_{\rm GRI}$ is the instant the GRI is encountered, its exact value depends on the accretion rate, e.g., for an accretion rate of $0.1$~\Msun/yr it can be estimated as $t_{\rm GRI} \simeq 10^{4.5}$\Msun$/\dot{M}$ \citep[see][]{herrington2023modelling}. 
    
    After $t_{\rm GRI}$ the central BH is formed with an initial mass of $M_{\rm BH,i}$. The BH accretion, $\dot{M}_{\rm BH}$, rate can also be written  as a function of the QS mass, following of \cite{santarelli2026evolutionary}: 
    \begin{equation}
        \dot{M}_{\rm BH} = \frac{1 - \epsilon}{\epsilon}\,\frac{4\pi}{\kappa c}\,\alpha\,G_{\rm TOV}\,M\  ,
        \label{eq:mdot_bh}
    \end{equation}
    where $\epsilon = 0.1$ is the radiative efficiency, $\kappa$ is the opacity of the material surrounding the BH and set at the inner boundary of the model, $M$ is the total mass of the QS. The free parameter $\alpha$  is used to account for possible decreases or enhancements in $\dot{M}_{\rm BH}$ due to angular momentum transport or magnetic fields, respectively.
    
    For the QS we consider two cases: either the QS continues to accrete at the same rate as its progenitor, or it stops accreting completely. We define the end of the QS phase as the instant the central BH has accreted the remaining stellar envelope, i.e. $M_{\rm BH} = M$.

    The mass of the BH grows following Eq.~\ref{eq:mdot_bh}, 
    \begin{equation}
        \dot{M}_{\rm BH} =
        \begin{cases}
        \frac{2}{\tau'}\  M(t_{\rm GRI}) & \text{if}\ \dot{M}=0,\\
        \frac{2}{\tau'}\, \dot{M}\, t &\text{if}\ \dot{M}>0.
        \end{cases}
        \label{eq:mdot_bh_cases}
    \end{equation}
    Where $\tau' = \left( \frac{1-\epsilon}{\epsilon}\, \frac{2\pi}{\kappa c}\, \alpha\, G \right)^{-1}$, this quantity has units of time.
    
    After integrating over time from $t_{\rm GRI}$ until the end of the QS phase, i.e. $t_{\rm end}$, we find that the BH mass is
    \begin{equation}
        M_{\rm BH} =
        \begin{cases}
        M_{\rm BH,i} + \frac{2}{\tau'}\  M(t_{\rm GRI})\ (t_{\rm end} - t_{\rm GRI}) & \text{if}\ \dot{M}=0,\\
        M_{\rm BH,i} + \frac{1}{\tau'}\, \dot{M}\, (t_{\rm end} - t_{\rm GRI})^2 &\text{if}\ \dot{M}>0.
        \end{cases}
        \label{eq:m_bh_cases}
    \end{equation}

    At $t_{\rm end}$ we have $M_{\rm BH} / M = 1$. In the case where $\dot{M}=0$ we obtain the following asymptotic behavior for $t_{\rm end}$:
    \begin{equation}
        t_{\rm end,no\ acc} \sim
        \begin{cases}
        {\tau'}/{2}     & \text{if}\ t_{\rm GRI} \ll \tau',\\
        {3\tau'}/{2}    &\text{if}\ t_{\rm GRI} \rightarrow \tau',\\
        t_{\rm GRI}     & \text{if}\ \tau' \ll t_{\rm GRI}.
        \end{cases}
        \label{eq:tend_noacc}
    \end{equation}

    For the case where the QS keeps accreting, its lifetime is determined as:
    \begin{equation}
        t_{\rm end,acc} \sim
        \begin{cases}
        {\tau'}      & \text{if}\ t_{\rm GRI} \ll \tau',\\
        \frac{3+\sqrt{5}}{2} \ {\tau'}   &\text{if}\ t_{\rm GRI} \rightarrow \tau',\\
        t_{\rm GRI}  + \sqrt{t_{\rm GRI}\ \tau'}   & \text{if}\ \tau' \ll t_{\rm GRI}.
        \end{cases}
        \label{eq:tend_acc}
    \end{equation}
    In all cases $t_{\rm end,acc} > t_{\rm end,no\ acc}$, and 
    \begin{equation}
        \frac{t_{\rm end,acc}}{t_{\rm end,no\ acc}} \sim
        \begin{cases}
        2 &  \text{if}\ t_{\rm GRI} \ll \tau',\\
        1 &\text{if}\ \tau' \ll t_{\rm GRI}.
        \end{cases}
        \label{eq:frac_qs_lifetime}
    \end{equation}

    As the lifetime of QSs depends mainly on $\tau'$, we estimate its value as it depends only on $\epsilon$, $\alpha$ and $\kappa$. If $\epsilon = 0.1$, $\alpha=1$ and $\kappa = 0.38$ cm$^2$/g (this value is extracted from our simulations and is mostly constant during the QS phase), one obtains $\tau' \simeq 10^8$ yr. This value is consistent with the estimates from our simulations. Also, in Appendix~\ref{app:relevant_quant} one can notice that the estimated lifetime of the accreting QSs is approximately twice as the non-accreting ones, consistent with  Eq.~\ref{eq:frac_qs_lifetime}.

    Lastly, we compare the the lifetime of the QS progenitor to the duration of the QS phase as
    \begin{equation}
        \frac{t_{\rm QS\ prog.}}{t_{\rm QS}} = \frac{t_{\rm GRI}}{t_{\rm end} - t_{\rm GRI}}\ .
    \end{equation}
    For an accretion rate of $0.1$~\Msun/yr it corresponds to $t_{\rm GRI}\sim 10^5$~yr. As $t_{\rm end}\sim 10^8$~yr, then we estimate, for this particular case,  that the QS phase lasts one thousand times more than the life of its progenitor. 
    If the accretion rate increases, then $t_{\rm GRI}$ decreases and vice versa \citep[see][]{herrington2023modelling}, while $t_{\rm end}$ remains mostly constant.

It is important to note that the estimate for $t_{\rm end}$ should be considered an upper limit for the QS lifetime. Compared with our simulations, the value of $\dot{M}_{\rm BH}$ is higher than the one obtained from Eq.~\ref{eq:mdot_bh}, which being more consistent with $\alpha>1$. As well, our results on $t_{\rm end}$ depend on the assumption that the QS can exist until $M_{\rm BH} / M = 1$, but the maximum BH mass that can support an envelope in hydrostatic equilibrium remains uncertain, with previous estimates going from $~0.1\%$ up to $\sim 60\%$ of the QS mass \citep[see][]{2011MNRAS.414.2751B,2012MNRAS.421.2713B,2024ApJ...970..158C,2026ApJ...998...65H,santarelli2026evolutionary}. In the case the QS end its lifetime when  $M_{\rm BH} / M = f_{\rm max}$, where $0< f_{\rm max}\le1 $, the resulting QS lifetime will be on the order of the product $f_{\rm max}\, t_{\rm end}$ as given by Eq.~\ref{eq:tend_acc} and \ref{eq:tend_noacc} for cases where $t_{\rm GRI} \rightarrow \tau'$ or $t_{\rm GRI} \ll \tau'$.

\section{Mass, luminosity and effective temperature across the stellar evolution and QS phases}
\label{app:relevant_quant}

    Table~\ref{tab:results} shows the evolution of stellar mass, luminosity, effective temperature and BH mass for three different epochs of the models' evolution, in the cases where the QS keeps accreting mass or not. First, such quantities are reported for the instant when the GRI condition is met, at that time the BH mass is zero. The second epoch corresponds to the point when the simulations crash. The last section shown the extrapolated values for the end of life of the QS and the resulting BH mass for different values of the parameter $\alpha$ from Eq.~\ref{eq:mdot_bh}.

    \begin{sidewaystable}[h!]
    \centering
    \setlength{\tabcolsep}{3.5pt}
    \renewcommand{\arraystretch}{1.2}
    \caption{Relevant quantities for the models with maximum mass gain rates $\dot{M}_{\rm gain,max} = 0.1$ and 1 ~\Msun/yr, $Y=0.25$ and $Z=10^{-4}$. }

    \begin{tabular}{cc|cccc|ccccc|cccc|ccc}

    $\dot{M}_{\rm gain,max}$ & Acc.& $t_{\rm GRI}$ & $M_{\rm GRI}$ & $L_{\rm GRI}$ & $T_{\rm eff,GRI}$ 
    & $t_{\rm crash}$ & $M_{\rm crash}$ & $L_{\rm crash}$ & $T_{\rm eff,crash}$ & $M_{\rm BH,crash}$ 
    & 
    & \multicolumn{3}{c|}{$t_{\mathrm{end,extrap}}$ } 
    & \multicolumn{3}{c}{$M_{\mathrm{BH,extrap}}$ } \\

    & QS&  &  &  &  
    &  &  &  &  &  
    & 
    &  &[Myr]  & 
    &  & [$10^6$~\Msun]&  \\

    [\Msun/yr]& &[Myr] & [$10^6$~\Msun] & [$10^9$~\Lsun] & [K] 
    & [Myr] & [$10^6$~\Msun] & [$10^9$~\Lsun] & [K] & [\Msun]
    & $\alpha =$
    & 0.5 & 1 & 1.5
    & 0.5 & 1 & 1.5 \\
    
    \hline

    0.1 & Yes & 0.349 & 0.035 & 1.291 & 7220 & 
                2.218 & 0.256 & 11.255 & 4542 & 10174 & & 
                187.46 & 93.952 & 62.074 &
                18.634 & 9.36 & 6.126 \\

    0.1 & No & 0.349 & 0.035 & 1.291 & 7220 & 
               3.275 & 0.035 & 2.284 & 4554 & 3099 & &
               88.148 & 45.208 & 31.276 &
               0.035 & 0.035 & 0.035 \\

    \hline
    1 & Yes &  0.072 & 0.068 & 2.461 & 7928 & 
               2.245 & 2.228 & 222.031 & 4856 & 110167 & &
               178.294 & 89.359 & 59.039 &
               176.644 & 88.739 & 58.084 \\

    1 & No & 0.072 & 0.068 & 2.461 & 7928 & 
             2.74  & 0.068 & 4.814 & 4588 & 5791 & & 
             88.672 & 45.476 & 30.746 &
             0.068 & 0.068 & 0.068 \\

      \hline

    \end{tabular}
    \label{tab:results}
    \end{sidewaystable}

    \section{Abundance patterns for of Na, Al and Mg}
    \label{app:mgal_abundances}

\begin{figure}[h!]
       \centering
        \includegraphics[width=0.8\columnwidth]{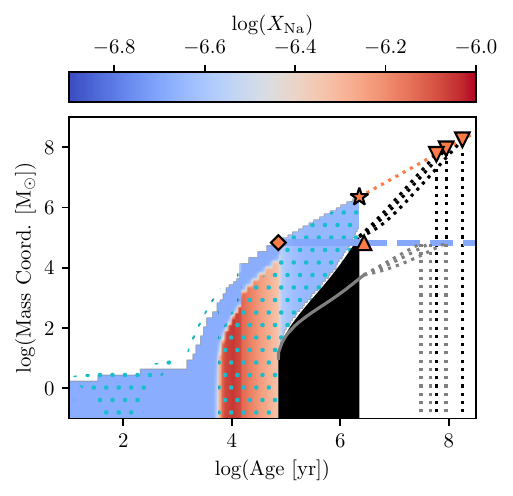}   \includegraphics[width=0.8\columnwidth]{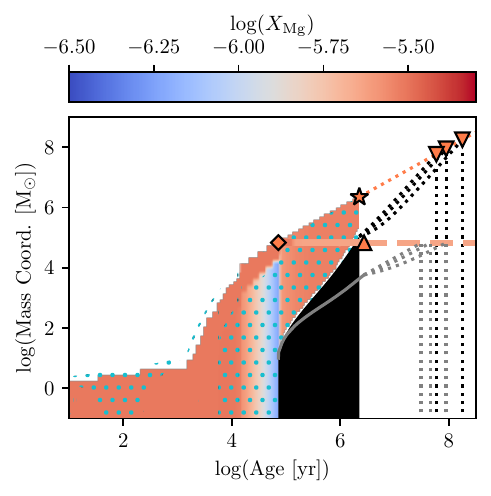} 
        \includegraphics[width=0.8\columnwidth]{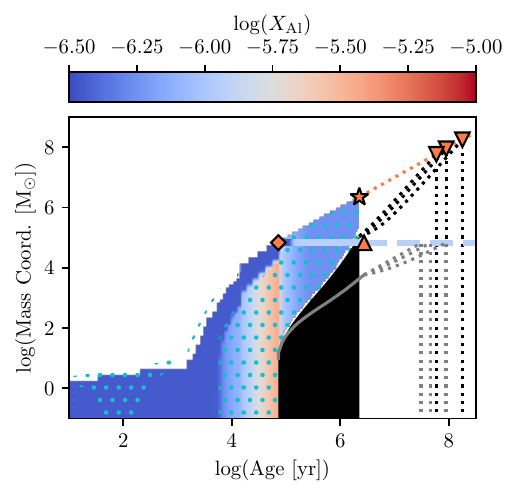} 
        \caption{Same as Fig.~\ref{fig:kipp_1} but with the mass fraction of Na (top), Mg (middle), and Al (bottom) being color-coded. In each panel the color of the thick dotted line corresponds to the constant surface abundance of Na, Mg, and Al respectively in the  case of the non-accreting QS, the  mass of which  stays constant.}
     %\vspace*{-1cm}
     \label{fig:kipp_1_NaAlMg}
    \end{figure}

    Fig.~\ref{fig:kipp_1_NaAlMg} shows Kippenhahn diagrams for the model with $\dot{M}_{\rm gain,max} = 1$~\Msun/yr, $Z=10^{-4}$, and $Y=0.25$, with sodium, magnesium, and aluminum mass fractions in different colors. In the first panel, sodium remains unchanged at early times, is enhanced in the core once the star reaches $\sim 100$~\Msun\ and the central temperature is $ > 40$~MK, and is then depleted for $> 70$~MK; during the QS phase, the surface Na abundance relaxes back toward its initial value as the remaining envelope becomes fully convective and the material is diluted with the unprocessed material. The other panels show the behavior for magnesium and aluminum.

\end{appendix}

\end{document}